\documentclass[aps,prl,twocolumn,superscriptaddress]{revtex4}
\usepackage{amssymb}
\usepackage{graphicx}

\begin{document}
\newcommand{\BSCCO}{Bi$_2$Sr$_2$CaCu$_2$O$_{8+\delta}$ }
\newcommand{\tc}{T$_{\mathrm c}$ }
\newcommand{\ar}{$A(\vec{r},\omega)$ }
\newcommand{\ak}{$A(\vec{k},\omega)$ }
\newcommand{\aq}{$A(\vec{q},\omega)$ }
\newcommand{\ef}{E$_{\mathrm F}$ }
\newcommand{\pn}{P$_{\mathrm N}$ }

\title{Elastic Scattering Susceptibility of the High Temperature Superconductor \BSCCO: A Comparison between Real and Momentum Space Photoemission Spectroscopies}

\author{K. McElroy}
\affiliation{Material Sciences Division, Lawrence Berkeley
National Lab., Berkeley, CA 94720 USA} \email[]{KPMcElroy@lbl.gov}
\author{G.-H. Gweon}
\affiliation{Physics Department, University of California
Berkeley, CA 94720 USA}
\author{S. Y. Zhou}
\affiliation{Physics Department, University of California Berkeley,
CA 94720 USA}
\author{J. Graf}
\affiliation{Material Sciences Division, Lawrence Berkeley National
Lab., Berkeley, CA 94720 USA}
\author{S. Uchida}
\affiliation{Department of Physics, University of Tokyo, Yayoi,
2-11-16 Bunkyoku, Tokyo 113-8656, Japan}
\author{H. Eisaki}
\affiliation{AIST, 1-1-1 Central 2, Umezono, Tsukuba, Ibaraki,
305-8568 Japan.}
\author{H. Takagi}\affiliation {Department of Advanced Materials Science, University of Tokyo, Kashiwa, Chiba 277-8561, Japan}
\affiliation {CREST, Japan Science and Technology Agency, Saitama
332-0012, Japan} \affiliation {RIKEN (The Institute of Physical
and Chemical Research), Wako 351-0198, Japan}
\author{T. Sasagawa}
\affiliation {Department of Advanced Materials Science, University
of Tokyo, Kashiwa, Chiba 277-8561, Japan} \affiliation {CREST,
Japan Science and Technology Agency, Saitama 332-0012, Japan}
\author{D.-H. Lee}
\affiliation{Material Sciences Division, Lawrence Berkeley
National Lab., Berkeley, CA 94720 USA}
\affiliation{Physics Department, University of California Berkeley,
CA 94720 USA}
\author{A. Lanzara}
\affiliation{Material Sciences Division, Lawrence Berkeley National
Lab., Berkeley, CA 94720 USA}
\affiliation{Physics Department, University of California Berkeley,
CA 94720 USA}

\date{\today}

\begin{abstract}
The joint density of states (JDOS) of \BSCCO is calculated by
evaluating the autocorrelation of the single particle spectral
function \ak measured from angle resolved photoemission
spectroscopy (ARPES). These results are compared with Fourier
transformed (FT) conductance modulations measured by scanning
tunneling microscopy (STM). Good agreement between the two
experimental probes is found for two different doping values
examined. In addition, by comparing the FT-STM results to the
autocorrelated ARPES spectra with different photon polarization,
new insight on the form of the STM matrix elements is obtained.
This shines new light on unsolved mysteries in the tunneling data.
\end{abstract}

 \maketitle

Atomically resolved spectroscopic studies taken by scanning
tunneling microscopy (STM) have given great insights into the low
energy electronic structure of high T$_{\mathrm c}$
superconductors. In particular, recent work using Fourier
transform STM (FT-STM) to study modulations in the local density
of states
(LDOS)\cite{hoffman02,howald03,mcelroy03,mcelroy04,mcelroy05} have
generated a great deal of
interest\cite{kivelson03,wang02,capriotti03,podolsky03}. When
these modulations are analyzed within a simple quasiparticle
interference model, motivated by the joint density of states
(JDOS)\cite{wang02}, a bridge between the momentum space probe
angle resolved photoemission spectroscopy (ARPES) and the real
space probe STM has been
established.\cite{mcelroy03,mcelroy04,mcelroy05} This connection
proves that STM and ARPES are consistent measurements of the same
underlying physical phenomenon. In addition, it allows for the two
diametrically opposite spectroscopic probes to compliment each
other.

Recently, R. Markiewicz \cite{markiewicz04} proposed a way to
check the validity of the JDOS model by comparing the FT-STM
result to the autocorrelation between the ARPES spectra
(AC-ARPES). In this Letter we report on such an analysis. In
addition to providing additional support to the JDOS
picture\cite{hoffman02, mcelroy03, wang02}, we find that the
autocorrelated ARPES spectra undergoes significant variation as
the photon polarization is changed. We propose that by comparing
them with the FT-STM result, we learn valuable information about
the STM matrix element. If our proposal is correct, some of the
long standing puzzles in the STM data can be resolved. These
include the lack of zero-bias peak in the tunneling spectra in the
vortex core\cite{vortex}, and the saturation of the quasiparticle
interference wave vectors near the Fermi energy
(\ef$\!\!\!$)\cite{mcelroy03,kivelson03}.

We begin by briefly discussing what STM and ARPES measure. First,
STM measures the tunneling current between an atomically sharp tip
and a clean flat surface at an energy $\omega$, relative to the
\ef, and at a location $\vec{r}$. Within the independent tunneling
approximation differential conductance, $g(\vec{r},\omega)$, is
given by:
\begin{equation}
g(\vec{r},\omega)=I_{0}|M_{f,i}^{\vec{r}}|^{2}R(\omega)A({\vec{r},\omega})\label{STM}
\end{equation}
where $M_{f,i}^{\vec{r}}$ is the tunneling matrix element,
$R(\omega)$ is due to the finite resolution of the STM, and \ar is
the real space single particle spectral function, commonly
referred to as the ``LDOS''\cite{weisendanger}. On the other hand,
ARPES measures the photo current $I(\hat{e},\vec{k},\omega)$ of
electrons ejected from a surface by UV and x-ray photons. In the
sudden approximation the photo current is given by:
\begin{equation}
I(\hat{e},\vec{k},\omega)=I_{0}|M_{f,i}^{\hat{e}}|^{2}f(\omega)A({\vec{k},\omega})\label{ARPESeq}
\end{equation}
where $M_{f,i}^{\hat{e}}$ are the photoemission matrix elements
(which depends on, among other things, the photon polarization
$\hat{e}$), $f(\omega)$ is the Fermi function, and \ak is the
momentum space single particle spectral function\cite{hufner}.

To check the JDOS picture\cite{markiewicz04} one needs to calculate
the JDOS directly from ARPES measured \ak by taking the following
autocorrelation
\begin{equation}
\mathrm{JDOS}(\vec{q},\omega)\!\! = \int\!\!\!
A({\vec{k}+\vec{q},\omega})A({\vec{k},\omega})\ \ d^{2}k.
\label{mark}
\end{equation}
Thus, barring matrix element effects (which we will come back to),
the AC-ARPES gives the susceptibility of a system to scattering
within the JDOS model.

Here we report such a AC-ARPES study on the high temperature
superconductor \BSCCO for two different doping values in the
superconducting regime\cite{Chatterjee05}.  When compared, the
AC-ARPES results and the corresponding FT-STM data show good
agreement for energy smaller than the superconducting gap. At
higher energies, additional weak features along the $(\pi,\pi)$
direction, not reported by FT-STM, are observed.  Most importantly
by changing the photon polarization from the nodal ($\Gamma$-X) to
the antinodal ($\Gamma$-M)) direction we suppress the contribution
from the nodal quasiparticles in the AC-ARPES. Interestingly, when
this is done, the AC-ARPES agrees better with the FT-STM data. We
propose that this supports the STM matrix element having a similar
suppression of the nodal states\cite{Wu00,balatsky}.

The experiments were carried out at beam line 10.0.1 of the
Advanced Light Source of the Lawrence Berkeley National Laboratory
and beam line 5.4 of the Stanford Synchrotron Radiation
Laboratory. Data from two representative \BSCCO samples are
presented. Both samples were measured with $0.15^{\circ}$ angular
resolution in the superconducting state at 25 K. The underdoped (UD64K)
sample, \tc of 64 K, was measured with polarization $\hat{e}\parallel\Gamma$M,
$h\nu=22$ eV and energy resolution of 22 meV. The
optimally doped sample (OPT92K), \tc of 92 K, was measured with $\hat{e}\parallel\Gamma$X,
$h\nu=33$ eV and resolution of 15 meV.

\begin{figure}[t!]
\includegraphics{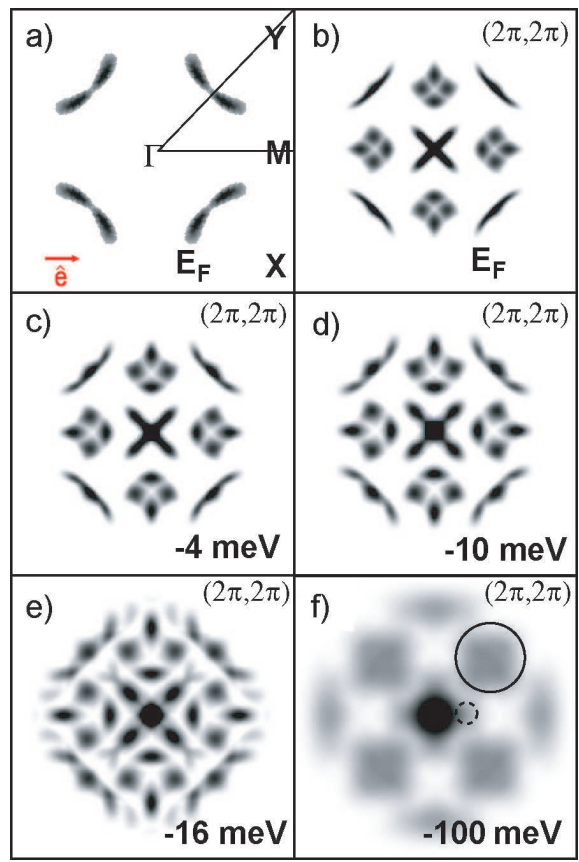}
\caption{\label{ARPES}a) ARPES photo current intensity measured in
the octant (indicated by the triangle) and then reflected to fill
in the first Brillouin zone for the UD64K sample. The red arrow
shows the polarization of the incoming photons relative to the
crystal structure. b)-e) Autocorrelations of the ARPES intensity
at \ef$\!\!$,-4,-10,-16 meV respectively. 32 different peaks can
be seen that disperse in a systematic fashion as a function of
energy. The same color scale is used in all the panels where black
represent maxima of intensity and white zero intensity. f)
Autocorrelation of ARPES intensity at -100 meV (outside the
superconducting gap). Weak features along $\vec{q}\ \|\ (\pi,\pi)$
(solid circle) can still be seen. Additionally no well defined
feature at $\vec{q}\approx (0,2\pi/4.5a_0)$ is seen at this energy
(dashed circle).}
\end{figure}

Figure \ref{ARPES}a shows a map of the ARPES spectral intensity at
\ef for the UD64K sample. Data were taken in the irreducible first
Brillouin zone and symmetrized. The high intensity points (black)
form typical arcs (banana shaped contours) near the nodal
direction and gap of spectral intensity as we approach the
antinodal point M, as expected for a d-wave superconductor. For
clarity, we have used a low intensity cutoff (less than 1/3 of the
maximum value) to suppress the contribution from the first and
second order superstructure replicas; this cutoff has no other
effect on the results discussed in this Letter. In this case an
additional suppression of spectral intensity is evident along the
nodal direction which is caused by the ARPES matrix element
\cite{asensio03} associated with the incoming photon polarization,
$\hat{e}\ \|\ \Gamma$M. In panels b)-e) we show the autocorrelated
ARPES spectra\cite{autocorrelation_nofold} for several energies
below \ef$\!\!$. The maps exhibit distinctive patterns of peaks,
whose $\vec{q}$-vectors disperse as the binding energy increases.
The qualitative features shown in Fig. \ref{ARPES} are robust
across different dopings; however, they become less distinct in
the overdoped regime.

\begin{figure}[t!]
\includegraphics{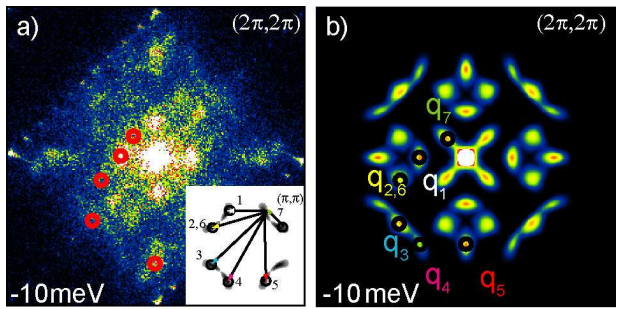}
\caption{\label{identification} a) FT-STM intensity at -10 meV
adapted from \cite{mcelroy03}. Red circles mark the peaks
associated with the `octet' model. b) A representative AC-ARPES
result at -10 meV with same peaks identified. The scattering
processes and designations that result in the `octet' peaks seen
in these two panels are shown in the inset in panel a).}
\end{figure}

In order to characterize these peaks, as well as compare them with
the FT-STM results, we show, in Figure \ref{identification}, a
direct comparison between the FT-STM  map (panel
a)\cite{mcelroy04} and AC-ARPES map (panel b) for a similar doping
and energy. All peaks in the AC-ARPES map are symmetry related to
the set of circles which are labeled using the notation used in
Ref.\cite{mcelroy03}. Each of the $q_i$ are the wave vectors
transfer associated with the elastic scattering between an `octet'
of momenta\cite{mcelroy03} (see inset in Fig. \ref{identification}
panel a)). Good agreement between the two spectroscopies is
observed in the sense that not only are all of the patterns
present in the AC-ARPES map also observed in the FT-STM map, but
they appear to occur at similar momentum locations. Additionally,
even the shapes of these patterns agrees very well between the two
maps. For example, the orientations of the ``cigar'' shaped $q_1$
and $q_5$ are similar in both maps with the four $q_1$'s making a
square while the four $q_5$'s make a cross. In the case of the
AC-ARPES we can clearly also distinguish $q_4$ which is not easily
identified by FT-STM \cite{mcelroy03}.

This analysis can be extended to energies outside the
superconducting gap energy. In particular, an additional q vector
$\vec{q}^{*}\approx(2\pi/4.5a_0,0)$ has been reported between
binding energies of 65 and 150 meV in underdoped samples and
interpreted as evidence of a competing charge order
parameter\cite{mcelroy05}. Fig. \ref{ARPES} f) shows the AC-ARPES
for a similar doping and energy (-100 meV). No distinct feature is
visible near $\vec{q}^{*}$ (dashed circle Fig. \ref{ARPES} f)) in
the AC-ARPES analysis. The fact that $\vec{q}^{*}$ does not show
up in the AC-ARPES suggests that it is not a quasiparticle
interference feature supporting the interpretation that it is due
to enhanced scattering from a competing order
parameter\cite{mcelroy05, ghosal04}. Also, at -100 meV the
AC-ARPES shows a weak, previously unreported feature along
$\vec{q}\|(\pi, \pi)$ (solid circle) in Fig. \ref{ARPES} f). We
believe this feature is due to the flat band located around
$(\pm\pi, 0)$ and $(0,\pm\pi)$. It will be interesting if this
feature is also seen by FT-STM.

\begin{figure}[t!]
\includegraphics{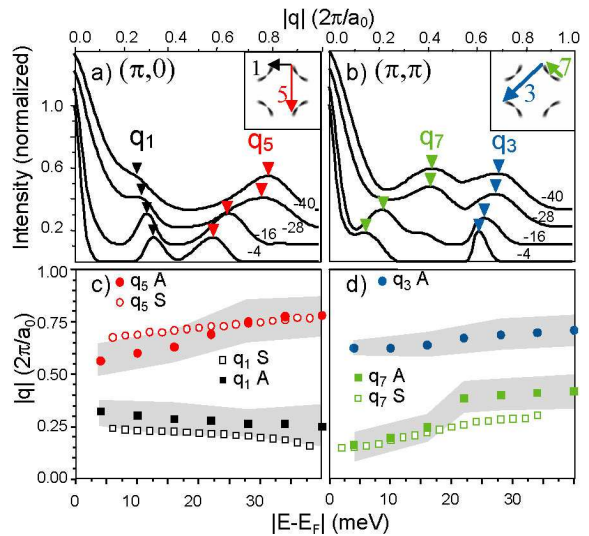}
\caption{\label{dispersions}a) Line cut parallel to $(\pi,0)$
through the AC-ARPES map for several energies (-4, -16, -28,
-40meV) near the \ef$\!\!$, for the UD64K sample. For clarity each
curve's maximum was normalized to unity. Dispersing peaks are
indicated by triangles. b) Linecut parallel to $(\pi,\pi)$ through
the AC-ARPES map for the same energies as in panel a. c-d) Peak
locations from AC-ARPES (full symbols), identified with the
corresponding q-vectors from the octet model, as a function of
energy for peaks in the $(\pi,0)$ (panel c) and $(\pi,\pi)$ (panel
d) direction. Corresponding STM peaks from \cite{mcelroy05} are
also plotted (empty symbols) for a similar doping. The gray areas
represent the FWHM of the peaks. No published FT-STM data exists
for q$_3$ at a similar doping.}
\end{figure}

To get a quantitative comparison between these results, we show in
Fig. \ref{dispersions} cuts through the AC-ARPES data along high
symmetry axes, for $q\ \|\ (\pi,0)$ (panel a) and $q\ \|\
(\pi,\pi)$ (panel b). By following the local maxima of these peaks
as a function of energy we have extracted the dispersion relation
for each of the vectors. The resulting $\vec{q}$-vector
dispersions are plotted in Fig. \ref{dispersions} c) and d)
(filled symbols) and are compared with the FT-STM results
\cite{mcelroy05} for a similar sample (open symbols).

We note here that, within a quasiparticle scattering theory, there
are other, off-shell, terms that contribute to the LDOS. However,
the excellent agreement between the two probes shown here suggests
that equation \ref{mark} is the dominant contribution to the
spatial-varying LDOS. Although this is in agreement with
single-impurity T-matrix calculations\cite{wang02}, the physical
reason is unknown.

The above successful comparison between AC-ARPES and the FT-STM
suggests the validity of the JDOS picture in the quasiparticle
interference model\cite{mcelroy03,hoffman02,wang02,mcelroy05}. Now
we turn to a main point of this Letter - the photon polarization
dependence of the agreement between the AC-ARPES and the FT-STM.
Until now we have swept under the rug the fact that the ARPES
intensity is not simply \ak (Eqn. \ref{ARPESeq}). Aside from the
Fermi function (which has little effect at low temperature and
non-zero binding energy), the ARPES spectra contain a matrix
element which depends on the polarization of the incident photon.
Thus far, we have used the data with photon polarization along
$\Gamma$M. Such photon polarization suppresses the states in the
$\Gamma$Y direction\cite{asensio03}. In the literature a similar
suppression of nodal states in the STM spectra has been suggested
to exist. This has been used to explain the spatial pattern of the
impurity states \cite{balatsky}, and the lack of a zero-bias
conductance peak in the vortex core\cite{Wu00}. Until now,
however, investigating the tunneling matrix element has been
limited by the lack of control that experimenters have over the
tip wavefunction upon which it depends \cite{weisendanger}.

\begin{figure}[t!]
\includegraphics{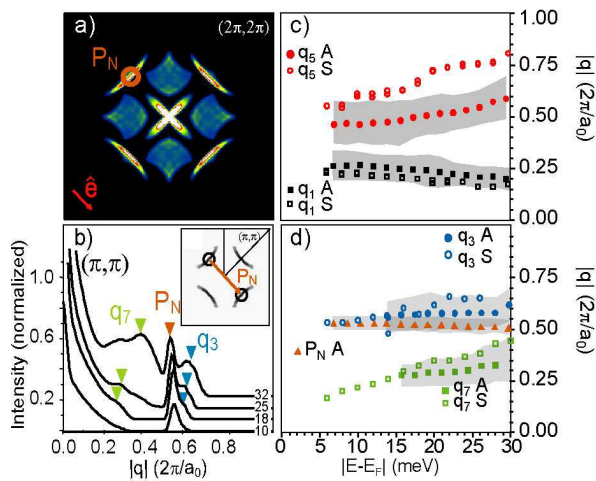}
\caption{\label{polarization}a) AC-ARPES measured on the OPT92K
crystal at -16 meV, in the triangle indicated, and with the photon
polarization parallel to the $\Gamma$X\ direction (red arrow). One
striking difference from Fig. \ref{ARPES} is the strong intensity
along $(\pi,\pi)$. b) Normalized cuts along the $(\pi,\pi)$
direction of the AC-ARPES intensity in a) (inset shows the origin
of \pn and the octant in which the data was taken). Clearly the
dominating feature is \pn$\!\!\!$. Small features can still be
seen at the `octet' wavevectors. c) `Octet' wavevectors
dispersions vs. energy along $\vec{q}\|(\pi,0)$. d) Dispersions of
the `octet' wavevectors and \pn feature along
$\vec{q}\|(\pi,\pi)$.}
\end{figure}

Figure \ref{polarization} illustrates the effect of changing the
photon polarization on our comparison. In panel \ref{polarization}
a) we show the AC-ARPES obtained for photons with polarization
parallel to the $\Gamma$X \ direction (red arrow) for the 92KOPT
sample. In this geometry the intensity in the FS map (see inset of
Fig. \ref{polarization} b) is enhanced along the nodal direction
with respect to the previous case (see Fig. \ref{ARPES}). The
AC-ARPES results in this case are dominated by an additional
bright spot, \pn (circled) arising from scattering between the
nodes. This vector can be also well resolved above the gap energy,
up to about 400 meV and disperses as the inverse of twice the
nodal Fermi velocity as expected for node-node scattering. Cuts
along the $(\pi,\pi)$ direction are shown in panel b), where the
\pn peak can be clearly observed, with respect to Fig.
\ref{dispersions} c). The dispersion of this new scattering vector
is plotted in panel d) and its location and dispersion makes it
easy to identify it with the node-node scattering vector. The
features associated with the `octet' model are still evident in
the AC-ARPES results for this photon polarization. Fig.
\ref{polarization} c) and d) show the dispersion of these
$\vec{q}$-vectors as a function of energy below the gap value
compared with the FT-STM results at the same doping. The agreement
with FT-STM observed for this polarization is worse (see e.g.
$\vec{q}_5$) than the one obtained in Fig. \ref{dispersions}.

Based on these observations we propose that the better agreement
between the AC-ARPES and the FT-STM for $\hat{e}\parallel \Gamma$M
compared to $\hat{e}\parallel \Gamma$X is due to a suppression of
nodal states in the STM spectra resulting from the tunneling
matrix elements\cite{balatsky,Wu00}. In addition to explaining the
vortex core results\cite{Wu00}, this suppression strengthens JDOS
features used to motivate the `octet' model but without
qualitatively changing their form. Simple theoretical JDOS
calculations show that such a matrix element will indeed cause the
flattening of the dispersion near zero-bias in the FT-STM data
without the need for competing order
parameters.\cite{mcelroy03,kivelson03,hoffman02,mcelroy05} While
the energy resolution used here is still relevant for detailed
discussion of low energy states\cite{kivelson03}, the robust
effect of the ARPES matrix element supports our proposal's
primacy.

In conclusion, the AC-ARPES analysis \cite{markiewicz04} was used
to calculate the JDOS of \BSCCO for two different doping values,
and the results were compared to FT-STM\@. The good agreement
found between the two probes support the JDOS interpretation
\cite{wang02} used to explain the density of states modulation
observed in FT-STM \cite{hoffman02, mcelroy03}. In addition we
have shown that by changing the photon polarization, and therefore
the ARPES matrix element, we can gain new insights into the
tunneling matrix element and its role in masking nodal states in
STM experiments. This has implications concerning the vortex core
spectra and the flattening of quasiparticle interference
dispersions near \ef$\!\!\!$. Finally, AC-ARPES resolves weak
features along $\vec{q}\|(\pi,\pi)$ and outside the
superconducting gap so far not reported by FT-STM.

\begin{acknowledgments}
We acknowledge and thank A. V. Balatsky, M. R. Norman, U.
Chatterjee, A. Kaminskii for very helpful discussions and Mr. M.
Ishikado for sample growth. This work was supported by the Director,
Office of Science, Office of Basic Energy Sciences, Division of
Materials Sciences and Engineering of the U.S Department of Energy
under Contract No. DEAC03-76SF00098; by NSF Grant No. DMR03-49361
and DMR04-39768, and the Sloan Foundations.
\end{acknowledgments}


\begin{thebibliography}{99}
\bibitem{autocorrelation_nofold} Backfolding in AC-ARPES was eliminated by setting the intensity to zero outside the first Brillouin zone to eliminate Umklapp processes absent in FT-STM.
\end{thebibliography}
\end{document}